

\magnification=\magstep1
\hsize=16truecm
\vsize=23.5truecm
\font\titlerm=cmbx10  scaled \magstep2
\font\chaprm=cmbx10  scaled \magstep1
\font\secrm=cmbx12

\newcount\secno
\newcount\susecno
\newcount\sususecno
\newcount\equnumber
\newcount\notenumber
\secno=0
\notenumber=1
\def\section#1{\goodbreak\bigskip\medskip\global\advance\secno by 1
                  \susecno=0\equnumber=0
       \centerline{{\chaprm \the\secno .  #1}}\smallskip\noindent}
\def\subsection#1{\goodbreak\bigskip\global\advance\susecno by 1
                  \equnumber=0\sususecno=0
       \noindent{ {\secrm \the\secno .\the\susecno. #1}} \medskip\noindent}
\def\subsubsection#1{\goodbreak\bigskip\global\advance\sususecno by 1
       \noindent{ {\bf \the\secno .\the\susecno .\the\sususecno. #1}}
                 \smallskip}
\def\fnote#1{\footnote{$^{(\the\notenumber )}$}{#1}
                 \global\advance\notenumber by 1}
\equnumber=0
\def\eqa{\global\advance\equnumber by1\eqno( \the\secno.
\the\equnumber )}
\def\lie#1{ \, \ell _{ #1 }\,}
\def\Lie#1{ \hbox{{\it \$\/}} \! {}_{ #1 }\,}
\def\dual{{{} {}^*}\! }
\def\varabl#1#2{ {\delta \,{ #1 } \over \delta \, { #2 }} }
\def\parabl#1#2{ {\partial \,{ #1 } \over \partial \, { #2 }} }
\newcount\refnumber
\refnumber=0
\def\refn{\global\advance\refnumber by1{\the\refnumber }}
\def\newref{\smallskip
              \item{ {\bf \refn .}}}
\def\xo#1{ \buildrel {\circ} \over {#1} {} \!\! }
\def\bg#1{ \buildrel {\scriptscriptstyle b} \over {#1} {} \!\! }
\def\HE{ \buildrel {E} \over H }
\def\LT{ {}_{{}_{LL}} \!\! t }
\def\LU{ {}_{{}_{LL}} \!\! U }
\def\hor#1{ {\underline{#1} } }
\bigskip\bigskip
\centerline{\titlerm Mass and Spin of Poincar\'e Gauge Theory}
\bigskip\bigskip
\centerline{\chaprm by}
\bigskip
\centerline{{\chaprm Ralf D. Hecht}\footnote{${}^\star $}{\sl
Present address: Institute for Theoretical Physics,
University of Cologne, D-50923 K\"oln, Germany}}
\bigskip
\centerline{{\bf Department of Physics,
            National Central University,}}
\centerline{{\bf Chung-Li, Taiwan 32054, ROC}}
\bigskip\bigskip\bigskip\bigskip

\centerline{\chaprm Abstract}
\bigskip
We discuss two expressions for the conserved quantities (energy momentum and
angular momentum) of the Poincar\'e Gauge Theory. We show, that the variations
of the Hamiltonians, of which the expressions are the respective boundary
terms, are well defined, if we choose an appropriate phase space for
asymptotic flat gravitating systems. Furthermore, we compare the expressions
with others, known from the literature.

\vfill\eject

\section{ Introduction}

If one looks for expressions of energy for a gravitating system, a natural
candidate for such an expression will be given by the Hamiltonian.
In gravitational theories the Hamiltonian can be written in the form $H= dB+J
\cong dB$, where $J$ vanishes for exact solutions. Consequently the energy of
exact solutions may be defined by $E := \int _\Sigma H = \int _{\partial
\Sigma} B$, where $\Sigma $ is a 3-dim. spacelike hypersurface. However the
Hamiltonian is not completely fixed by the requirement of generating the
correct field equations, it can be modified by adding a total divergence or
equivalently a boundary term at spatial infinity. As pointed out by Regge and
Teitelboim [1] in the case of General Relativity, one has to adjust the
boundary term in such a way that the variation of the Hamiltonian is well
defined; this means that no variations of the derivatives of the variables
occur. But this argumentation of Regge and Teitelboim fixes only the
integrals, not the integrands, and the whole discussion depends on the phase
space choosen. Therefore some freedom in constructing energy expressions still
exists.

In this paper, we will discuss two possible boundary terms for energy
momentum and angular momentum of the Poincar\'e Gauge Theory (PGT). One of
the expressions was given by Nester [2], the other one (see [3]) is a
modification of it. Both expressions were tested in [4] with exact solutions,
but a detailed discussion has not been given.
We will show that the variations of the respective Hamiltonians  are well
defined, in the sense of Regge and Teitelboim, if we choose an appropriate
phase space for asymptotic flat gravitating systems.

Suitable expressions for the conserved quantities of the PGT for asymptotic
flat solutions were given earlier by Hayashi and Shirafuji [5] and by
Blagojevi\'c and Vasili\`c [6]. In their works they have to restrict themself
to an asymptotic Cartesian basis. Also approaches were made for calculating
conserved quantitites of the PGT in asymptotic anti-de Sitter space times,
see [7] and [8], but they didn't proof to be successful.
One advantage of the expressions discussed here is that they need no
restriction to an asymptotic Cartesian basis and can be evaluated also in
asymptotic anti-de Sitter space times.

First we will give a brief introduction into the framework of the PGT. Then we
will calculate the fall off of asymptotic flat solutions of the PGT in order
to be able to fix the phase space. In section 3 we will write down the
Hamiltonian and the expressions we will deal with. The variation of the
Hamiltonian and the argumentation of Regge and Teitelboim are worked out in
section 4. In section 5 we show that the integrals of our expressions are
indeed finite and conserved. Finally, in section 6, we will compare them with
the work of Hayashi \& Shirafuji [5] and Blagojevi\'c \& Vasili\`c [6].

\bigskip

Let us shortly recapitulate the underlying theory and fix the conventions.
The PGT (see, for instance, [9,10]) is a gauge theory of gravity
in which spacetime is represented by a 4-dimensional Riemann-Cartan
manifold. The gauge potentials are the orthonormal basis 1-forms
$\vartheta^{  \alpha } $ and the connection 1-forms $\omega _{  \alpha }
{}^{  \beta } $. The corresponding field strengths are the torsion
$ \Theta ^{  \alpha } = D \vartheta ^{  \alpha } := d \vartheta ^{  \alpha }
+ \omega _{  \mu } {}^{  \alpha } \wedge \vartheta ^{  \mu } $ and the
curvature $ \Omega _{  \alpha } {}^{  \beta } := d \omega _{  \alpha }
{}^{  \beta } + \omega _{  \gamma }{}^{  \beta } \wedge \omega _{  \alpha }
{}^{  \gamma }$.
The sources of the gravitational fields
are the 3-forms of material energy-momentum $\Sigma _\alpha $ and spin
angular-momentum $\tau _\alpha {}^\beta $ which are variational derivatives
of the material Lagrangian with respect to the gauge potentials.
In order to have a local Poincar\'e invariant Lagrangian for the
gravitational field, the gravitational Lagrangian should be of the form
$$ L=L_G(\vartheta ^\alpha , \Theta ^\alpha , \Omega _\alpha {}^\beta )
   + L_M(\vartheta ^\alpha , \psi, D\psi ) \quad . \eqa $$
Variation with respect to the potentials yield the field equations
$$ DH_\alpha -\epsilon _\alpha  = \Sigma _\alpha  \quad  \hbox{and}  \quad
   DH_{\alpha \beta } - \epsilon _{\alpha \beta } = \tau _{\alpha \beta }
   \quad , \eqa $$
where
$$ H_{\alpha} := -
   {{\partial L_G}\over{\partial d\vartheta^{\alpha}}} =
   - {{\partial L_G}\over{\partial \Theta ^{\alpha}}}\qquad \quad , \quad
   \quad  H^{\alpha}{}_{\beta} := - {{\partial L_G}\over{\partial
   d\omega_{\alpha}{}^{\beta}}} =
   - {{\partial L_G}\over{\partial
   \Omega _{\alpha}{}^{\beta}}} \quad , \eqa$$
and
$$ \epsilon _\alpha = e_\alpha  \rfloor L _G + (e _\alpha  \rfloor \Theta ^\mu
)
   \wedge H_\mu  + ( e_\alpha \rfloor \Omega _\mu  {}^\nu  )\wedge
   H ^\mu {}_\nu \quad , \quad
   \epsilon _ {\alpha \beta } = \vartheta _{[\beta } \wedge H_{\alpha ]}
   \quad . \eqa $$
In this article we restrict ourself to Lagrangians, which are at most
quadratic in the field strengths. This leads to
$$ L= {\Lambda _{cos} \over l^2} \eta - {1 \over 2} \Theta ^\alpha  \wedge
   H_\alpha -{1 \over 2} \Omega _\alpha {}^\beta \wedge H^\alpha {}_\beta +
   {a_0 \over 4l^2} \Omega _\alpha {}^\beta \wedge \eta ^\alpha {}_\beta
   \quad . \eqa $$
where $ \Lambda _{cos} $ is the cosmological constant, $l $ the Planck
length, $\eta $ the volume 4-form,  $\eta ^{{  \alpha } {  \beta } }=
{}^*( \vartheta ^{  \alpha } \wedge \vartheta ^{  \beta } )$, and ${}^ * $
the Hodge star. Then the field momenta can be expressed in terms of the
irreducible pieces of the field strength [11]:
$$ H_\alpha=-{1\over l_0^2 }{}{ *\atop {} } \left(
   \sum_{n=1}^3  a_n \, {}^{(n)} \Theta _\alpha \right) \quad , \quad
   H_{\alpha \beta} = -{a_0\over 2l_0^2 } \eta_{\alpha \beta } \;
   -\; {1\over \kappa }{}{ *\atop {} }\left( \sum_{n=1}^6
   b_n \, {}^{(n)} \Omega _{\alpha \beta} \right) \quad , \eqa $$
where $\kappa $ , $a_i $, and $b_i$ are coupling constants. As we are
interested in asymptotic flat solutions, we set $\Lambda _{cos} =0$.

In this article we use Greek letters to denote anholonomic indices,
and Latin letters for holonomic indices. The metric is given by $g_{\alpha
\beta} = diag(-1,1,1,1)$.

We will need some technical details: The connection $\omega _\alpha {}^
\beta = r_\alpha {}^\beta + K_\alpha {}^\beta$ splits into a purely
Riemannian part $r_\alpha {}^\beta $ and the contortion $K_\alpha {}^\beta$.
The purely Riemannian part of the curvature is denoted by $R_\alpha {}^\beta $
(Riemann 2-form). The Lie derivative of a scalar valued form $\Psi $
with respect to a vector field $\xi $ is given by $\lie \xi \Psi := \xi
\rfloor d \Psi + d (\xi \rfloor \Psi)$. For tensor valued forms we have to use
$$ \Lie v \Psi _\alpha {}^\beta := \lie v \Psi _\alpha {}^\beta
   + \Psi _\mu {}^\beta l _\alpha {}^\mu  - \Psi _\alpha {}^\mu l _\mu
   {}^\beta \qquad , \quad
   l_\alpha {}^\beta := e_\alpha \rfloor \lie \xi \vartheta ^\beta \quad,
   \eqa $$
which may be more generally written as
$$ \Lie v \Psi = \lie v \Psi - l_\alpha {}^\beta S^\alpha {}_\beta \Psi
   \quad , \eqa $$
where $S^\alpha {}_\beta $ is the generator of the Lorentz group in the
respective representation.

\section{The fall off of asymptotic flat PGT solutions}

For our discussion of the Hamiltonian and its boundary term we have to choose
the phase space of our system. Therefore we consider first the behaviour of
asymptotically flat exact solutions of the PGT (compare [9],[12]).

We demand the solutions to be asymptotically  flat and we use an asymptotically
Cartesian coordinate system. Therefore we have a radial coordinate $r$. For a
function, we define the fall off by
$$  \lim _{r \to \infty} (r^n f) =constant  \; \Longleftrightarrow : \;
    f=O_n  $$
etc.. We say a $p-$form $\omega $ is $O_n$, $\omega = O_n$, iff all their
components with respect to the asymptotically Cartesian basis are at least
$O_n$.

We start with the requirement
$$ e _i {}^\alpha \longrightarrow \delta _i{}^\alpha + O_1 \quad , \quad
   e _i {}^\alpha {}_{,j} \longrightarrow O_2 \quad \hbox{and } \quad
   \omega _{i \alpha }{}^\beta  \longrightarrow O_1 \quad , \quad
   \omega _{i \alpha }{}^\beta{}_{,j}  \longrightarrow O_2 \quad . \eqa $$
where $e_i{}^\alpha $ are the components of the basis 1-forms and
$\omega _{i\alpha}{}^\beta $ the components of the connection forms with
respect to the asymptotically Cartesian holonomic basis: $\vartheta ^\alpha =
e_i{}^\alpha dx^i$, $\omega _\alpha {}^\beta =\omega _{i\alpha}{}^\beta dx^i$.
Of course, it follows that $ g_{ij} = o_{ij} + O_1$, $g_{ij,k} = O_2 $
($o_{ij}$ is the Minkowskian metric tensor).
For the field strength we have $ \Theta^\alpha = O_1$, $d \Theta^\alpha =O_2 $,
$ \Omega _\alpha {}^\beta = O_2$, $d \Omega _\alpha {}^\beta = O_3$.

As we are interested in the asymptotics of isolated gravitating systems,
we only consider the vacuum field equations.
For our purpose it is useful to split the momenta into an ``Einsteinian''
part and the rest
\fnote{If we choose the paramater $a_i,b_i$ such that $H_{\alpha \beta} =
\eta _{\alpha \beta} /(2l^2)$ and $H_\alpha =0$, then we get just Einsteins
theory, whereas in the teleparallelism ($\Omega _\alpha {}^\beta =0$) the
choice $H_\alpha = -K^{\mu\nu}\wedge \eta_{\mu\nu\alpha}/(2l^2)$ leads to
the teleparallel equivalent of General Relativity.
}:
$$ H_\alpha =:\bar H_\alpha  -a_0{\buildrel {E} \over  H }_\alpha
   \quad \hbox{where }  {\buildrel {E} \over  H }_\alpha :=
   - {1 \over 2l^2} K^{\mu \nu }\wedge \eta _{\mu \nu \alpha } \quad \hbox{and}
   \eqa $$
$$ H_{\alpha \beta } =: -{a_0 \over 2l^2} \eta _{\alpha \beta }
   + \bar H _{\alpha \beta } =:  \bar H _{\alpha \beta } - a_0
   {\buildrel {E} \over  H }_{\alpha \beta } \quad . \eqa $$
Then $\bar H _\alpha {}^\beta $ is at least $O_2$. Because of $ D{\buildrel
{E} \over  H }_{\alpha \beta } \equiv  \vartheta _{[\alpha} \wedge
{\buildrel {E} \over  H }_{\beta ]}$ we get from the second field equation
$ D{ \bar H }_{\alpha \beta } =  \vartheta _{[\alpha} \wedge \bar
H _{\beta ]}$, which can be solved for $\bar H_\alpha $:
$$ \bar H_\alpha =
   e^\mu \rfloor D\bar H _{\alpha \mu } + {1 \over 4} \vartheta _\alpha
   \wedge ( e^\nu \rfloor e^\mu \rfloor D \bar H _{\mu \nu } ) \quad , \eqa $$
and the irreducible decomposition of this equation gives
$$ (a_n + a_0 {\buildrel {\scriptscriptstyle E} \over {a}} {}_n) {}^{(n)}
   \Theta ^\alpha  = O_3 \quad ,\eqa $$
where $[{\buildrel {\scriptscriptstyle E} \over {a}} {}_1,
{\buildrel {\scriptscriptstyle E} \over {a}} {}_2,
{\buildrel {\scriptscriptstyle E} \over {a}} {}_3]=[-1,2,{1 \over 2} ]$.
Because of the identity
$$ D\HE _\alpha - {1\over 2} (e_\alpha \rfloor \Theta ^\beta ) \wedge
   \HE _\beta + {1\over 2} \Theta ^\beta \wedge (e_\alpha \rfloor \HE _\beta )
   \equiv {1\over 2 l^2 } (R ^{\mu \nu } - \Omega ^{\mu \nu } )\wedge
   \eta _{\mu \nu \alpha } \quad ,\eqa $$
$(\eta ^{\alpha \beta \gamma} =\dual (\vartheta ^\alpha
\wedge \vartheta ^\beta \wedge \vartheta ^\gamma ) $)
we get in the case that $\bar H_\alpha =0$, for the left hand side of the
first field equation (which we will call $F_\alpha$ ):
$$ F_\alpha  = {a_0 \over 2 l^2} (\Omega _\mu {}^\nu - R _\mu {}^\nu ) \wedge
   \eta ^\mu {}_{\nu \alpha } - {a_0 \over 2l^2} \Omega _\mu {}^\nu \wedge
   \eta ^\mu {}_{\nu \alpha } + \hbox{ squares of curvature } \; .  \eqa $$
Here the first term originates from the torsion part, and the last two terms
from the curvature part of the field equation.
If $\bar H _\alpha $ is not equal to zero, then it is of order $O_3$
(see (1.6,2.3)) and gives no new contribution to the last equation
(up to $O_4$ terms). Hence, in any case we have
$$ F_\alpha  = O_4 - {a_0 \over 2 l^2} R _\mu {}^\nu \wedge  \eta ^\mu {}_
   {\nu \alpha } \quad . \eqa $$
If we require $\omega _\alpha {}^\beta = O_{1+\gamma }$ with $\gamma > 0$, then
the term $O_4$ in the equation (2.7) above will change into $O_{4+\gamma }$,
and also $\bar H _\alpha $ will be of order $O_{3+\gamma}$.

\section {The Hamiltonian}

We rewrite the Lagrangian in the following form (see [13])
$$ L = - \Theta ^\alpha \wedge H_\alpha - \Omega _\alpha {}^\beta \wedge
   H^\alpha {}_\beta + D \psi \wedge P - \Lambda (\vartheta ^\alpha ,
   H_\alpha ,H^\alpha {}_\beta ,\psi , P) \quad . \eqa $$
The field equations follow from the variational principle regarding the
potentials $\vartheta ^\alpha $, $\omega _\alpha {}^\beta $, $\psi $ and the
momenta $H_\alpha $, $H^\alpha {}_\beta $ and $P$ as independent.
The potential $\Lambda $ is quadratic in $H_\alpha$ and $H^\alpha
{}_\beta$ in such a way, as to reproduce relations equivalent to (1.2).
The Lagrangian is invariant under diffeomorphisms and $SO(3,1)$ rotations.
This invariance leads in the usual way to the Noether identities.
We vary the Lagrangian,
$$ \delta L = d \left[  - \delta \vartheta ^\alpha \wedge H_\alpha -
   \delta \omega _\alpha {}^\beta \wedge H^\alpha {}_\beta
   + \delta \psi \wedge P \right]
   + \delta \vartheta ^\alpha \wedge \varabl L {\vartheta ^\alpha }
   + \delta \omega _\alpha {}^\beta \wedge \varabl L {\omega _\alpha
   {}^\beta } + $$
$$ \delta \psi \wedge \varabl L \psi +
   \varabl L {H_\alpha } \wedge  \delta H_\alpha
   + \varabl L {H^\alpha {}_\beta } \wedge  \delta H ^\alpha {}_\beta
   + \varabl L P \wedge \delta P  \eqa $$
and deal only with the symmetry transformations Lorentz rotations and
diffeomorphisms. Then we have for instance $\delta \vartheta ^\alpha =
\varepsilon ^\alpha {}_\beta \vartheta ^\beta - \lie \xi \vartheta ^\alpha $
or, $\delta \psi = \varepsilon ^\beta {}_\alpha S^\alpha {}_\beta
\psi - \lie \xi \psi $ etc. ($S^\alpha {}_\beta $ are the generators of
Lorentz rotations, $ \varepsilon ^\beta {}_\alpha $ arbitrary parameters, and
$\xi $ is an arbitrary vector field, generating the diffeomorphism).
Considering only diffeomorphism invariance, we will get eventually
the first Noether identity (compare [10], [13]):
$$ (\xi \rfloor \vartheta ^\alpha )\wedge D\varabl L { \vartheta ^\alpha }
   + (-1)^{p+1}\, (\xi \rfloor \psi ) \wedge D \varabl L \psi
   + D \varabl L {H_\alpha } \wedge (\xi \rfloor H_\alpha)
   +D \varabl L {H^\alpha {}_\beta } \wedge (\xi \rfloor H^\alpha {}_\beta )
   + $$
$$ (-1)^{p+1} D \varabl L P \wedge (\xi \rfloor P) =
   (\xi \rfloor D \vartheta ^\alpha )\wedge \varabl L {\vartheta ^\alpha }
   + (\xi \rfloor \Omega _\alpha {}^\beta )\wedge \varabl L
   {\omega _\alpha {}^\beta }
   + (\xi \rfloor D\psi ) \wedge  \varabl L \psi + $$
$$ \varabl L {H_\alpha } \wedge (\xi \rfloor D H_\alpha )
   + \varabl L {H^\alpha {}_\beta } \wedge (\xi \rfloor D H^\alpha {}_\beta )
   + \varabl L P \wedge (\xi \rfloor D P) \quad , \eqa $$
and we get the second Noether identity by using the Lorentz invariance
of the Lagrangian,
$$ D \varabl L {\omega _\alpha {}^\beta} = - \vartheta
   {}^{[\alpha }\wedge \varabl L {\vartheta ^{\beta ]}} - S^\alpha {}_\beta
   \psi \wedge \varabl L \psi + \varabl L {H_{[\alpha } }\wedge
   H_{\beta ]} - \varabl L {H^\beta {}_\mu} \wedge H^\alpha {} _\mu $$
$$ + \varabl L {H^\mu {}_\alpha } \wedge H^\mu {}_\beta +
   \varabl L P \wedge P S^\alpha {}_\beta \quad . \eqa $$

The invariance of the Lagrangian leads also to the Noether current.
We can identify the Noether current 3-form from (3.2) as
$$ H = \xi \rfloor L - \delta \vartheta ^\alpha \wedge H_\alpha -
   \delta \omega _\alpha {}^\beta \wedge H^\alpha {}_\beta +
   \delta \psi \wedge P \quad . \eqa $$

For a timelike vector field $\xi $ and vanishing $\varepsilon $, the Noether
current is just the canonical Hamiltonian of the theory, and therefore we
will call in future $H$ also the (generalized) Hamiltonian.

This Hamiltonian can be recast in the form:
$$ H \equiv  dB + J := d \left[  \xi ^\alpha \wedge H_\alpha  + \tilde
   \varepsilon ^\beta {}_\alpha \wedge H^\alpha {}_\beta - (\xi \rfloor \psi )
   \wedge P \right] +
   (\xi \rfloor \vartheta ^\alpha ) \wedge \varabl L {\vartheta ^\alpha } +
   \qquad \qquad \qquad $$
$$ (\xi \rfloor \psi) \wedge \varabl L \psi +
   \tilde \varepsilon ^\beta {}_\alpha \wedge \varabl L
   {\omega_\alpha {}^\beta} + \varabl L {H_\alpha}\wedge (\xi\rfloor H_\alpha )
   + \varabl L {H^\alpha {}_\beta  } \wedge (\xi \rfloor H^\alpha{}_\beta )
   + (-1)^{p+1} \varabl L P \wedge (\xi \rfloor P) \quad , \eqa $$
where $\tilde \varepsilon ^\beta {}_\alpha := \varepsilon ^\beta {}_\alpha +
\xi \rfloor \omega _\alpha {}^\beta $. Obviously, the Hamiltonian is weakly
conserved (that is conserved for exact solutions), $dH \cong 0$.

For a space time symmetry (a Killing field $\xi $) it is known that the
gravitational part of the Hamiltonian is conserved, even if the gravitational
field equations are not fulfilled. With the help of the Noether identities,
we calculate the derivative of the Hamiltonian ($\tilde \omega $ is the
transposed connection
\fnote{The name is only appropriate, if one chooses a holonomic basis.
},
$\tilde \omega _\alpha {}^\beta := \omega _\alpha {}^\beta + e_\alpha \rfloor
\Theta ^\beta $):
$$ dH = dJ =
   \Lie \xi \omega _\alpha {}^\beta \wedge \varabl L {\omega _\alpha {}^\beta}
   + \Lie \xi \psi \wedge \varabl L \psi + \varabl L {H_\alpha }
   \wedge \Lie \xi H ^\alpha + \qquad \qquad  $$
$$ \qquad \qquad  \varabl L {H^\alpha {}_\beta } \wedge
   \Lie \xi H^\alpha {}_\beta  + \varabl L P \wedge \Lie \xi P +
   d\left[ ( \tilde \varepsilon ^\beta {}_\alpha - \tilde D_\alpha \xi ^\beta )
   \wedge \varabl L {\omega _\alpha {}^\beta } \right] \quad . \eqa $$

We see, that if $\xi $ is a Killing field and the matter field equation is
fulfilled, then the Hamiltonian is conserved, if we choose  $\varepsilon
^\beta {}_\alpha = l _\alpha {}^\beta$ (because of the identity $l _\alpha
{}^\beta + \xi \rfloor \omega _\alpha {}^\beta \equiv \tilde D _\alpha\xi ^
\beta $).

But we cannot simply use the boundary term $B$ of (3.6) as superpotential for
conserved quantitites. There exist mainly two obstacles:

One reason is that the variation principle, as used in PGT, doesn't give
a proper momentum (it leads just to $\eta _\alpha {}^\beta $) for the linear
(Hilbert-term) part of the Lagrangian. We can study the situation in the case
of GR:

 From the Hilbert Lagrangian we can get, by adding a total divergence, the
Lagrangian $L'$, which is constructed out of squares of the connection,
$$ L_H = -{1\over 2l^2} R_\alpha {}^\beta \wedge \eta ^\alpha {}_\beta =:
   L' - {1\over 2l^2}d( r_\alpha {}^\beta \wedge \eta ^\alpha {}_\beta )
   \qquad =: L' + d K  .\eqa $$
We choose $ L'$ as it does not contain second derivatives of the basis 1-form.
The variation of this Lagrangian is
$$ \delta L'  = \delta \vartheta ^\mu \wedge \left[ d { \buildrel {M} \over P
   } _\mu +{1\over 2 } d \vartheta _\alpha \wedge (e_\mu \rfloor { \buildrel
   {M} \over P } {}^\alpha ) -{1 \over 2} (e_\mu \rfloor d \vartheta _\alpha )
   \wedge { \buildrel {M} \over P } {}^\alpha \right] + d(\delta \vartheta ^
   \mu \wedge { \buildrel {M} \over P } _\mu )\quad ,   \eqa $$
the term in the brackets is the Einstein 3-form, and
the momentum is given by $ { \buildrel {M} \over P } _\alpha = -1/(2l^2) \,
r_\mu {}^\nu \wedge \eta ^\mu {}_{\nu\alpha } $ , which was first introduced
by M\o ller [16] and is a kind of anholonomic Freud potential. The Lagrangian
may be now rewritten as $L' \equiv 1/2 \, d \vartheta ^\alpha \wedge
{ \buildrel {M} \over P } _\alpha $.
We split the variation into a Lie derivative $ \delta _\xi = - \lie \xi $
and  a rotational part, where the generators of Lorentz transformations
are denoted by $\varepsilon ^\alpha {}_ \beta $ ( $\delta _\varepsilon
\vartheta ^\alpha = \varepsilon ^\alpha{}_\mu \vartheta ^\mu $), $\delta =
\delta _\xi + \delta _\varepsilon$. Rewriting (3.9) gives

$$ 0 = \delta \vartheta ^\alpha \wedge {\delta L \over \delta \vartheta
   ^\alpha } + d[ \delta \vartheta ^\alpha \wedge { \buildrel {M} \over P }
   {}^\alpha + \xi \rfloor L' + \delta _\varepsilon K]\quad .  $$
We identify the term in the brackets as the (generalized Hamiltonian or)
Noether current. For the variations we get
$$ \delta _\varepsilon \vartheta ^\alpha \wedge { \buildrel {M} \over P }
   _\alpha +\delta _\varepsilon K  = {1\over 2l^2} D(\varepsilon ^\beta
   {}_\alpha \wedge \eta ^\alpha {}_\beta ) \eqa $$
and
$$ - \lie \xi \vartheta ^\alpha \wedge { \buildrel {M} \over P } _\alpha +
   \xi \rfloor L' =
   - d(\xi ^\alpha { \buildrel {M} \over P } _\alpha ) + \xi ^\alpha \wedge
   {\delta L \over \delta \vartheta ^\alpha } \quad , \eqa $$
therefore the superpotential for the Noether current is
$$ - \xi ^\alpha { \buildrel {M} \over P } _\alpha + {1\over 2l^2}
   \varepsilon ^\beta {}_\alpha \wedge \eta ^\alpha {} _\beta \quad . \eqa $$
Apparently this term is not contained in $B$ of (3.6).

The second reason is, that the boundary term $B$ transforms
not homogeneously. This restricts the range of application of this term
to asymptotically Cartesian bases. If we want to improve $B$ in this respect,
we have to introduce an additional structure, a background field for instance.
Moreover, in general spacetimes there is no Killing field at our disposel.
But if we deal with spacetimes, which possess asymptotical symmetries,
it is natural to use these asymptotical symmetries in order to fix the free
parameters $\xi $ and $\varepsilon$. Therefore, we introduce a background
spacetime $\bg U _4$ which is a copy of the asymptotic regions (spacelike
infinity) of our physical spacetime $U_4$. This background space time
will allow us also to construct covariant expressions for the boundary
term, see below. The geometric quantities $\bg \vartheta ^\alpha$ and
$\bg \omega _\alpha {}^\beta $ of the background space time should not
exhibit any dynamics. In order to deal with both the physical and the
background quantities, we map the background space-time onto the physical
space-time by some diffeomorphism $f$ (we need only to identify the outer
regions of the space times) $f:\; \bg U _4 \longrightarrow U_4$. This induces
the forms $ \xo \vartheta ^\alpha :=f^{-1}{}^* \bg \vartheta {}^\alpha$
and $ \xo \omega _\alpha {}^\beta  :=f^{-1}{}^* \bg \omega _\alpha {}^\beta $
and vector fields $\xi := f_* \bg \xi $. For the diffeomorphism $f$ we demand
that
$$ g_{ij} = \bg g _{ij} + O(1/r) \eqa $$
for an appropriate coordinate system.
The mapping can be constructed by identifying the coordinate systems
$x$ of $\bg U _4$ and $y$ of $U_4$, with $  f= y^{-1} \circ x $
if the coordinate systems fulfill (3.14) (see, for instance, [14] for a
similar construction).

If we now vary the potentials and momenta of the physical spacetime,
the quantities $\bg \vartheta ^\alpha $ and $\bg \omega _\alpha {}^\beta $
remain of course fixed. Also $\xo \vartheta ^\alpha $ and $\xo \omega _\alpha
{}^\beta $ remain fixed, if the function $f$ does not change.
Notice that in this construction a change of $\xi $, induced by variations
of the potentials and momenta, can only occur, if the function $f$ is
affected by this variation. But in this case, also the quantities
$ \xo \omega _\alpha {}^\beta $ and $\xo \vartheta ^\alpha$ have to change.
We will not consider this possibility.

The Hamiltonian is not fixed by the conservation law or the field equations,
we can always add a surface term, $H=dB + J \longrightarrow H_i=dB_i + J$.
This freedom we will use to improve $B$ concerning the flaws
mentioned above and make the variation of the (improved) Hamiltonian
well defined in the sense of Regge and Teitelboim [1] (see below).
As we are only interested in the behavior of the boundary terms in the
asymptotic region of spacetime, we will henceforth neglect the matter fields,
which are supposed to vanish in this region.
Here we will show, that both of the following boundary terms
$$ B_1 = (\xi \rfloor \vartheta ^\alpha )\wedge \Delta H_\alpha  + \Delta
   \omega_\alpha {}^\beta \wedge (\xi \rfloor H^\alpha {}_\beta ) +
   (\xi \rfloor \omega _\alpha {}^\beta ) \wedge \Delta H_\alpha {}^\beta
   \quad , \eqa $$
(where $\Delta \omega _\alpha {}^\beta = \omega _\alpha {}^\beta
- \xo \omega _\alpha {}^\beta $ etc.) which was given by Nester [2] and (see
[3])
$$ B_2 = (\xi \rfloor \vartheta ^\alpha )\wedge \Delta H_\alpha  + \Delta
   \omega _\alpha {}^\beta
   \wedge (\xi \rfloor H^\alpha {}_\beta ) + \xo {\tilde D }_\alpha \xi ^\beta
   \wedge \Delta H^\alpha {}_\beta   \quad  \eqa $$
make the variations of the corresponding Hamiltonians well defined, if we
choose suitable phase spaces for asymptotic flat solutions. The expressions
were not deduced as Noether current of a suitable Lagrangian (like $B$), but
the improvements were done by hand and we will justify this choice later
(section 4). Observe that the expressions contains
a term $\omega _\alpha {}^\beta \xi \rfloor H^\alpha {}_\beta $, which equals
the M\o ller potential in leading order. The expression
$\varepsilon _\alpha {}^\beta = e_\alpha \rfloor \xo {\tilde D} \xi ^\beta $
which we choose in $B_2$, is a covariant generalization of the
generators of rotations in Minkowski space and is in fact antisymmetric, since
$\xi $ is a killing vector of the background.

The conserved quantities (total momentum and angular momentum) of
asymptotic flat solutions are now calculated by integrating the surface
term over a 2-sphere with radius $R$ and take the limes $R \rightarrow
\infty$:
$$ Q(\xi) := \lim_{R\to \infty} \int _{S^2} B(\xi) \eqa $$

By choosing the vector field $\xi $ to be one of the Killing-fields
of the Minkowski-space, one get the corresponding conserved quantity.
This calculations were done in [4] for both expressions with asymptotic flat
and asymptotic constant curvature solutions of PGT. The results were, for the
tested solutions, the same as for the corresponding solutions of General
Relativity.

\section{The variation of the Hamiltonian}

The variation of the  Hamiltonian has the general pattern:

$$ \delta \int _\Sigma H =  \int _\Sigma \left( \delta \vartheta ^\alpha
   \wedge a_\alpha + \delta \omega _\alpha {}^\beta  \wedge b^\alpha {}_\beta +
   c^\alpha \wedge \delta H_\alpha +  d_\alpha {}^\beta \wedge
   \delta H_\alpha {}^\beta + dX \right) \quad , \eqa $$
As pointed out by Regge and Teitelboim [1], we have to make sure that
$ \int _\Sigma d X$ vanishes, otherwise the field equations are not
expressible as variational derivatives of the Hamiltonian with respect
to the potentials and momenta. To reach this goal, we are free to add a
boundary term to the Hamiltonian.

Nester set $\varepsilon_\alpha {}^\beta =0$ and obtained the following term
from the variation of his Hamiltonian $H_1= dB_1+J$:
$$ X_1 =\delta ( \xi \rfloor \vartheta ^\alpha )\wedge \Delta H_\alpha
   - \delta \vartheta ^\alpha \wedge (\xi \rfloor H_\alpha )
   + \Delta \omega _\alpha {}^\beta \wedge \delta
   (\xi \rfloor H^\alpha {}_\beta ) + \qquad\qquad$$
$$ \qquad\qquad\qquad\qquad\delta (\xi \rfloor
   \omega _\alpha {}^\beta ) \wedge \Delta H^\alpha {}_\beta
   \quad . \eqa $$
The variation of the Hamiltonian $H_2=dB_2+J$ yields:
$$ X_2 = \delta (\xi \rfloor \vartheta ^\alpha ) \wedge \Delta H_\alpha
   - (\delta \vartheta ^\alpha ) \wedge (\xi \rfloor H_\alpha )
   + \Delta \omega _\alpha {}^\beta \wedge \delta (\xi \rfloor H^\alpha {}_
   \beta ) $$
$$ \qquad \qquad \qquad - (\xi \rfloor \omega _\alpha {}^\beta )\wedge
   \delta H^\alpha {}_\beta + \delta ( e_\alpha \rfloor \xo {\tilde D} \xi ^
   \beta ) \wedge \Delta H^\alpha {}_\beta \quad .\eqa $$
In (4.3) we set $ \delta ( e_\alpha \rfloor \xo {\tilde D} \xi ^\beta ) =0$,
because we do not vary the background quantities. For completness we also
write down the variation of the canonical Hamiltonian (3.6):
$$ X= \delta (\xi \rfloor \vartheta ^\alpha ) \wedge H_\alpha
   - \delta \vartheta ^\alpha \wedge (\xi \rfloor H_\alpha )
   + \delta (\xi \rfloor \omega _\alpha {}^\beta ) \wedge H^\alpha {}_\beta
   - \delta \omega _\alpha {}^\beta \wedge (\xi \rfloor H^\alpha {}_\beta )
   \quad . \eqa $$

These formulas are for no use, if we do not have a phase space given.
Here we are interested in asymptotically flat solutions. For the
potentials and momenta we do not demand that they fulfill the field
equations, but we require that they possess the same asymptotic fall off
as asymptotically flat solutions of the field equations as worked out in
Sec. 2. For the background we choose the Minkowski space and a Cartesian
basis. Then we have vanishing $\xo H _\alpha$, $\xo \omega _\alpha {}^\beta$,
and the field momentum of rotation is reduced to
$\xo H ^\alpha {}_\beta = -(a_0 / 2l^2) \xo \eta ^\alpha {}_\beta $. For the
potentials we require the fall off as given in (2.1) and for the momenta
as well as for their variations we require
$\bar H _\alpha = O_3$, $\bar H^\alpha {}_\beta = O_2$.

Now we can start with the variation of the Hamiltonian for asymptotic
flat spacetimes as specified above.
We begin with the variation of the canonical Hamiltonian and first consider
only translations.
$$ X = \xi \rfloor \left[ \delta \vartheta ^\alpha \wedge H_\alpha +
   \delta \omega _\alpha {}^\beta \wedge H^\alpha {}_\beta \right] \quad .
   \eqa $$
The integral $ \int _\Sigma d X =  \int _{\partial \Sigma } X $ will vanish,
if $X$ fall off faster than $r^{-2}$. Because the variations of the potentials
are independent, each of the terms should fall off faster than $O_2$ in order
to make the variation of the Hamiltonian well defined. But solutions will
in general not have this fall off. Moreover, the Hamiltonian will not give a
reasonable energy momentum-expression in the case of GR. Therefore we turn to
the Hamiltonian $H_2$. To the canonical Hamiltonian (3.6) we add the following
surface term:
$$ - d\left[ \xi \rfloor ( \omega _\alpha {}^\beta \wedge H^\alpha {}_\beta )
   \right] \quad . \eqa $$
The improved Hamiltonian is then given by
$$ \tilde H= d\left[ (\xi \rfloor \vartheta  ^\alpha )\wedge H_\alpha +
   \omega _\alpha {}^\beta \wedge \xi \rfloor H^\alpha {}_\beta \right] +
   J \quad , \eqa $$
which is just the Hamiltonian $H_2$ for the case that $\xi $ is a translational
Killing field of the background Minkowski space time, because in this case
$ \xo {\tilde D }_\alpha \xi ^\beta $ vanishes.
The variation of this Hamiltonian leads to a boundary term
$$ X_2 = \xi \rfloor \left[ \delta \vartheta ^\alpha \wedge H_\alpha -
   \omega _\alpha {}^\beta \wedge \delta H^\alpha {}_\beta \right] . \eqa $$
If we now use the relations (2.2,3), we obtain
$$ X_2 = \xi \rfloor \left[ \delta \vartheta  ^\alpha \wedge \bar H _\alpha
   + {a_0 \over 2l^2} \delta \vartheta ^\mu \wedge r_\alpha {}^\beta
  \wedge \eta ^\alpha {} _{\beta \mu } - \omega _\alpha {}^\beta \wedge
  \delta \bar H^\alpha {}_\beta \right] = \xi O_3 . \eqa $$
Therefore the variation of the generator of the translations is well-defined.

For spacelike rotations, the Killing field of the background is
$$ \xi ^i = x_N \delta ^i_M -x_M \delta ^i_N \; ,\qquad
   \hbox{ M,N = 1,2,3} \quad \hbox{fixed }.  \eqa $$
Now $\varepsilon ^\beta {}_\alpha $ no longer vanishes. As
$ \xo {\tilde D }_\alpha \xi ^\beta $ is evaluated on the background,
its variation vanishes and the variation of the Hamiltonian $H_2$ is also
given by (4.9).

We can write $X= \xi \rfloor Y$ (for $Y=Y^\alpha \eta _\alpha $), and we are
only interested in the projection of this integrand on the 2-sphere $t$=const,
$r$=const. This projection can be written as $\xi ^a Y^0 dS_a \sim \xi ^a Y^0
x_a d \theta d \phi =0$. Therefore the variation of the generator of the
rotations is well defined.

For the boosts we need stronger restrictions (compare [6]). We demand
$$ \omega _\alpha {}^\beta =O_{(1+\gamma)} \qquad \hbox{where }
   \gamma >0 \quad . \eqa $$
Then the last term of the rhs of (4.9) is of order $O_{3+\gamma }$,
and, because of
$$ \xi ^i = x_N \delta ^i_0 -x_0 \delta ^i_N , \eqa $$
we have
$$ X_2 = x_N \partial _t \rfloor \left[ {a_0 \over 2l^2} \delta
   \vartheta ^\mu \wedge r_\alpha {}^\beta \wedge \eta ^\alpha {}
   _{\beta \mu }\right] + O_{2+\gamma} \quad . \eqa $$
Furthermore we have to require parity conditions (compare [1],[5],[6]),
$$ e_i{}^\alpha = \delta _i^\alpha  + {a_i{}^\alpha (\hbox{\bf n})
   \over r} + O_{(1+\zeta )} \qquad \zeta >0 \quad ,
   a_i{}^\alpha (\hbox{\bf n}) =a_i{}^\alpha (-\hbox{\bf n}) \quad . \eqa $$

Then the Levi Civita connection is odd in leading order
$(r_\alpha {}^\beta = \mu  _\alpha {}^\beta + O_{(2+\zeta )}$, $\mu _\alpha
{}^\beta = O_2$ and $\mu _\alpha {}^\beta (\hbox{\bf n})=-\mu _\alpha {}^\beta
(-\hbox{\bf n}) )$.We find

$$  x_N \partial _t \rfloor \left[ {a_0 \over 2l^2} \delta
   \vartheta ^\mu \wedge r_\alpha {}^\beta \wedge \eta ^\alpha {}
   _{\beta \mu }\right] = \chi ^a dS_a \quad , \eqa $$
where $ \chi ^a dS_a = - 2x_N \delta e_i{}^\mu r_{j\alpha }{}^\beta
( e^{[i}{}_\beta e^{j]}{}_\mu e^{a\alpha } + e^{[i}{}_\mu e^{j]\alpha }
e^{a\beta } + e^{[i|\alpha } e^{|j]}{}_\beta e^a{}_\mu )\eta_{0a} $.
We recognize that in leading order (that is $O_2$) $\chi ^a$ is an even
function. Because $dS_a $ is odd, the integral over a 2-sphere will vanish.
Therefore the variations of the boost generators are well defined.

\bigskip

The variation of Nester's Hamiltonian gives a boundary term
$$ X_1 = \xi \rfloor (\delta \vartheta ^\alpha \wedge H_\alpha )
   + \delta (\xi \rfloor\omega _\alpha{}^\beta ) \wedge \Delta H^\alpha{}_\beta
   + \omega _\alpha {}^\beta \wedge \delta (\xi \rfloor H^\alpha {}_\beta ),
   \eqa $$
which can be rewritten into
$$ X_1 = \xi \rfloor ( \delta \vartheta ^\alpha \wedge \bar H _\alpha )
   + \omega _\alpha {}^\beta \wedge (\xi \rfloor\delta\bar H ^\alpha {}_\beta )
   - {a_0 \over 2 l^2} \xi \rfloor (\delta \vartheta ^\gamma \wedge r _\alpha
   {}^\beta \wedge \eta ^\alpha {}_{\beta \gamma} ) $$
$$ + {a_0 \over 2l^2}
   \delta \vartheta ^\gamma \wedge (\xi \rfloor \omega _\alpha {}^\beta )
   \wedge \eta ^\alpha {}_{\beta \gamma } - {a_0 \over 2 l^2}
   \delta (\xi \rfloor \omega _\alpha {}^\beta ) \wedge \Delta \eta ^\alpha
   {}_\beta + (\xi \rfloor \omega _\alpha {}^\beta ) \wedge \bar H^\alpha
   {}_\beta .\eqa $$
We see that in the case of translations, $X_1$ will fall off faster than
$1/r^2$. In the case of rotations, we have to impose the parity conditions
$$ e_i{}^\alpha = \delta _i {}^\alpha + {a_i {}^\alpha \over r} + O_2 \qquad
   \hbox{, with }a_i{}^\alpha \hbox{ even,}  \eqa $$
and the stronger fall off of the connection (4.11).
Beside the stronger fall off condition, we have also to notice that
$B_2$ -- contrary to $B_1$ -- transforms inhomogeneously under Lorentz
transformations. Therefore the whole discussion of $B_2 $
is basis dependent.

\section{Conservation and finiteness}

Now we turn to the conservation and finiteness of the integrals. We now
require that the variables fulfill the field equations.
For this purpose we first observe that
$$ \omega _\alpha {}^\beta \wedge (\xi \rfloor H^\alpha {}_\beta )=
   a_0 \HE _\mu \xi ^\mu  - {a_0 \over 2l^2} \xi ^\mu r_\alpha {}^\beta
   \wedge \eta ^\alpha {}_{\beta \mu } + \omega _\alpha {}^\beta \wedge
   (\xi \rfloor \bar H^\alpha {}_\beta ) \quad , \eqa $$
where from it follows that
$$ \xi ^\gamma H_\gamma + \omega _\alpha {}^\beta \wedge (\xi \rfloor
   H^\alpha {}_\beta ) = - {a_0 \over 2l^2} \xi ^\mu r_\alpha {}^\beta
   \wedge \eta ^\alpha {}_{\beta \mu } + \xi ^\gamma \bar H _\gamma +
   \omega _\alpha {}^\beta \wedge (\xi \rfloor \bar H ^\alpha {}_\beta )
   = $$
$$ -{a_0 \over 2l^2} \xi ^\mu r_\alpha {}^\beta \wedge
   \eta ^\alpha {}_{\beta \mu } + \xi O_3 \quad . \eqa $$
For translations the boundary term $B_2$ gives just
$$ Q(\xi) = - {a_0\over 2l^2}\int _{S^2} r_\alpha {}^\beta  \wedge \xi
   \rfloor \eta ^\alpha {}_\beta  \quad . \eqa $$

We have $r_\alpha {}^\beta \wedge d \eta ^\alpha {}_{\beta \mu}= O_4$, and
for exact solutions of the (PGT) field equations we find $ (d r_\alpha
{}^\beta ) \wedge \eta ^\alpha {}_{\beta \mu} = - (r_\nu {}^\beta \wedge
r_\alpha  {}^\nu) \wedge \eta ^\alpha {}_{\beta \mu} + O_4 = O_4$ and
therefore also $d( r_\alpha {}^\beta \wedge \xi \rfloor \eta ^\alpha
{}_\beta )= O_4$ ($\xi ^\mu =constant +O_1$). Consequently their 4-momentum
is conserved.

Now we consider the rotations. For exact solutions it is
$$ H_2 = d \left[ \xi ^\alpha  H_\alpha  + \omega _\alpha {}^\beta
   \wedge (\xi \rfloor H^\alpha {}_\beta ) + \varepsilon ^\beta {}_\alpha
   H^\alpha {}_\beta \right] = $$
$$ - {a_0 \over 2l^2} d \left[ \xi ^\mu  r_\alpha {}^\beta \wedge
   \eta ^\alpha {}_{\beta \mu } + \varepsilon ^\beta {}_\alpha
   \eta ^\alpha {}_\beta  \right] + d\left[ \xi ^\mu  O_3 +
   \varepsilon ^\beta {}_\alpha O_2 \right] \quad . \eqa $$
Apparently the integration over the last term gives a finite result.
The first term is a kind of anholonomic version of the Landau-Lifshitz
expression, the latter one reads in the language of the exterior calculus,
$$ M^{ik} = \int_{{\cal S}} (x^i \LT ^k -x^k \LT ^i ) =
   \int_{S^2} (x^i \LU ^k - x^k \LU ^i ) \; + \; {1\over 2l^2} \int_{S^2}
   \sqrt{-g}\; \eta ^{ik} \quad . \eqa $$
Here
$$ \LU ^i = {1\over 2l^2 }\sqrt{-g}\,(r_m{}^n \wedge \eta ^m{}_n{}^i )
   \quad , \quad   \LT ^i = d \LU ^i\eqa $$
are the superpotential and the energy complex of Landau-Lifschitz [15]
respectively. With the help of the parity conditions (4.14) and the fall off
(4.11), the expression (5.4) can be shown to be finite:

For rotations, the Killing field is given by (4.10), where now $M,N \, \in
\{0,1,2,3\}$. Consequently we have $ \partial _i \xi ^j = o_{Ni} \delta ^j{}_M
- o_{Mi} \delta ^j{}_N = \varepsilon ^j{}_i $.
Therefore $ d \xi ^\beta  = \varepsilon ^\beta {}_\alpha \wedge
\vartheta ^\alpha + O_1^e$ (because  $\varepsilon ^\beta {}_\alpha$
is evaluated wirth respect to the background basis, which differs from
the physical one by $O_1^e$) where $O_n^o$ ($O_n^e$) means $X+O_{n+\zeta}$ and
$X$ is a odd (even) term of order $O_n$
\fnote{The term $ \xo {\tilde D }
_\alpha \xi ^\beta \wedge \xo H^\alpha {}_\beta  $ gives no
contribution to the integral, as it is easily seen, if we use the theorem
of Stokes and $d(\xo {\tilde D } _\alpha \xi ^\beta  \wedge \xo H^\alpha
{}_\beta ) =\xo {\tilde D } _\alpha \xi ^\beta  \wedge d \xo \eta ^\alpha
{}_\beta /(2l^2)=0 $. But the term $ \xo H^\alpha {}_\beta $ is important in
te case of asymptotic anti-de Sitter spacetimes, for instance.}.

Thus we have $d\xi ^\mu \wedge r_\alpha {}^\beta \wedge \eta ^\alpha {}
_{\beta \mu } = - \varepsilon ^\beta {}_\alpha d\eta ^\alpha {}_\beta +
O_1^e \wedge r_\alpha {}^\beta \wedge \eta ^\alpha {}_{\beta \mu } +
O_{1+\zeta } \wedge r_\alpha {}^\beta \wedge \eta ^\alpha {}_{\beta \mu}$,
and the integration of this term will cancel the integration of $d (\varepsilon
^\alpha {}_\beta \wedge \eta ^\alpha {}_\beta )$, because of $d
\varepsilon =0$ and
$r_\alpha {}^\beta =  O_2^o$. Now it remains to show that the integration
over $\xi ^\mu d(r_\alpha ^\beta \wedge \eta ^\alpha {}_{\beta \mu})$ yields
a finite value. Because of $G_\alpha = O_{4 +\gamma}$, we have $dr_\alpha
{}^\beta \wedge \eta ^\alpha {}_{\beta \mu} = O_4^e$. Furthermore, we get
$r_\alpha{}^\beta \wedge d\eta ^\alpha {}_{\beta \mu} = O_4^e$. Thus,
$\xi ^\mu d(r_\alpha {}^\beta \wedge \eta ^\alpha {}_{\beta \mu} ) = O_3^o$,
and the integral over this term is finite.
Observe, that the only troublesome terms in (5.4) was the Einstein part. The
term $(\xi ^\mu O_3 + \varepsilon O_2)$ behaves well, and, imposing (4.11),
will make the integral over this term vanish.

Now we have
$$ {d \over dt} \int _\Sigma H = \int_\Sigma \lie {\partial _t} H =
   \int _S {\partial _t} \rfloor H \eqa $$
because of $dH=0$.
\goodbreak
Therefore the charge is conserved, if
$ {\partial _t} \rfloor H  $ falls off faster than $r^{-2}$, and this is the
case for $H_2$ as shown above.

\bigskip

To get finite quantities for the boundary term $B_1$, which are also
conserved, we have to impose stronger restrictions. We require that $\omega$
fulfills (4.11), and from (3.14) and (5.2) we find, that the only
non-vanishing term in the boundary integral is
$$ -{a_0\over 2l^2} \left[ r_\alpha {}^\beta \wedge (\xi \rfloor
   \eta ^\alpha {}_\beta ) +
   (\xi \rfloor \omega _\alpha {}^\beta) \wedge \Delta \eta _\alpha {}^\beta
   \right] \quad . \eqa $$
Because the torsion, contained in the connection, is independent and cannot be
cancelled out, we have to demand $\Theta _\alpha = O_{2+\gamma}$ (where $
\gamma > 0 $ is necessary for conservation) or parity conditions on the
torsion. For the other term recall that $d(r_\alpha {}^\beta \wedge \eta
^\alpha {}_{\beta \mu} )= O_4^e$, thus $d(\xi ^\mu r_\alpha {}^\beta \wedge
\eta ^\alpha {} _{\beta \mu}) = d \xi ^\mu \wedge (r_\alpha {}^\beta \wedge
\eta ^\alpha {}_{\beta \mu } ) + \xi ^\mu d (r_\alpha \wedge \eta^\alpha
{}_{\beta \mu } ) $,
where the last term is of order $\xi O_4^e$. Thus we have only to estimate
$(d\xi ^ \mu) \wedge (r_\alpha {}^\beta \wedge \eta ^\alpha {}_{\beta \mu})$.
For $B_2$ this term was canceled in leading order by $d(\varepsilon ^
\beta {}_\alpha \wedge \eta ^\alpha {}_\beta )$. Here we have to demand
stronger parity conditions for the tetrads. With
$$ \omega _{i\alpha } {}^\beta = O_{2+\gamma} \qquad \hbox{and} \qquad
   e_i{}^\alpha = \delta _i {}^\alpha + {a_i {}^\alpha \over r} + O_{2+\gamma}
   \qquad \hbox{with }a_i{}^\alpha \hbox{ even,}  \eqa $$
the boundary term $B_1$ will give finite and conserved quantities
(where $\gamma > 0$ is only needed to show the conservation of the
quantities).

\section{Relationship to other expressions for conserved quantities of PGT}

We compare the potential $B_2$ with expressions as given in [5] and [6].
The first investigations about conserved quantities in PGT were done by
Hayashi and Shirafuji [5]. In a Lagrangian approach, they started with
a generator like (3.6) and substituted for the vector field $\xi $ just the
Killing fields of Minkowski spacetime in a Cartesian basis. That is,
$$ \int_{{\cal S}  } H = {\rm c}^i \int_{\partial{\cal S}  }P_i +
   {\rm b}_{ij} \int_{\partial{\cal S} } L^{ij} + {\rm d}_{\alpha \beta }
   \int_{\partial{\cal S} } S^{\alpha \beta } \quad , \eqa $$
where momentum, angular momentum, and spin are given by
$$ P_i  = \left(-\parabl {L_{HS}} {d \vartheta ^\alpha } \right) e_i{}^\alpha +
   \omega _{i \mu}{}^\nu \left( -\parabl {L _{HS}} {d \omega _\alpha {}^\beta
   } \right) \quad , \eqa $$
$$ L^{ij} = x_{[i} \left[ e_{j]}{}^\alpha \left(-\parabl {L_{HS}}
   {d \vartheta ^\alpha } \right) + \omega _{j] \alpha }{}^\beta
   \left( -\parabl {L_{HS}} {d \omega _\alpha {}^\beta } \right) \right]
   \quad , \eqa $$
$$ \hbox{and} \qquad S^{\alpha \beta } = H^{\alpha \beta } + {a_0 \over 2l^2}
   \xo \eta ^{\alpha \beta } .\eqa $$

In order to get reasonable results, they had to write the Hilbert part
of the curvature in terms of the torsion and a divergence. The reason is
already discussed in chapter 3:  As the first part of (6.7) is not contained
in the canonical Hamiltonian, Hayashi and Shirafuji
wrote the Einstein-Cartan term for the Lagrangian (1.5) in the following way
$$ - {a_0\over 2l^2} \Omega _\alpha {}^\beta \wedge \eta ^\alpha {}_\beta =
   - {a_0\over 2l^2} \omega _\mu {}^\beta \wedge \omega _\alpha {}^\mu \wedge
   \eta ^\alpha {}_\beta - {a_0\over 2l^2}\omega _\alpha {}^\beta \wedge
   d \eta ^\alpha {}_\beta - {a_0\over 2l^2}d (\omega _\alpha {}^\beta \wedge
   \eta ^\alpha {}_\beta ) \quad .\eqa $$
They discarded the exact term: $L_{HS} = L + a_0 /( 2l^2 ) d (\omega _\alpha {}
^\beta \wedge \eta ^\alpha {}_\beta )$, where $L$ is given in (1.5).
Then the translational field momenta $(-\partial {L _{HS}}/ d \vartheta
^\alpha )$ picks up an additional term $-(a_0/ 2l^2) \omega _\mu {}^\nu \wedge
\eta ^\mu {}_{\nu \alpha }$.
Therefore, the field momenta of Hayashi and Shirafuji are given by
$(-\partial {L _{HS}}/ d \vartheta ^\alpha ) = H _\alpha -{a_0\over 2l^2}
\omega _\mu {}^\nu \wedge \eta ^\mu {}_{\nu \alpha }$ and
$-{L_{HS}} / {d \omega _\alpha {}^\beta }= \bar H^\alpha {}_\beta $.

Energy momentum and angular momentum are singled out by an appropriate
choice of the parameters c,b,d in (6.1), where b and d has to be coupled
by $\hbox{b}_{ij}= \xo e _i{}^\alpha \xo e _j{}^\beta d_{\alpha \beta} $.
The integrated quantities of Hayashi and Shirafuji then coincide with ours,
provided the condition (4.11) is fulfilled.

To compare the expression (3.10) with the work of Blagojevi\'c and Vasili\'c
[6], we evaluate our expression in the framework of the Ricci calculus.
The boundary term  is then given by $( 1/2 {\cal H}_\alpha {}^{jk} \eta_{jk}:=
e H_\alpha  \; etc., e:= det (e_i{}^\alpha )$
$$ e B _2 = {1\over 2} \left[ \xi ^\alpha {\cal H}_\alpha {}^{jk}
   + \omega _{i\alpha}{}^\beta \xi ^i {\cal H} ^\alpha {}_\beta {}^{jk}
   + 2  \omega _{i\alpha } {}^\beta \xi ^{[j|} {\cal H} ^\alpha {}_\beta
   {}^{|k]i}{} + \varepsilon ^\beta {}_\alpha \Delta {\cal H}^\alpha {}_\beta
   {}^{jk} \right] \eta _{jk} \eqa $$
where $\varepsilon ^\beta {}_\alpha$ is given by $\xo {\tilde D }_\alpha \xi
^\beta $. For the energy ($\xi = \partial _0$, $\varepsilon ^\beta {}_\alpha =
0$) we obtain
$$ \hor B _2 =  {1\over e} \left[ {\cal H} _0 {}^{0a} - \omega _{b\alpha }
   {}^\beta {\cal H} ^\alpha {}_\beta {}^{ba} \right] dS_a \quad ,\eqa $$
where $\hor B $ means the projection onto the spacelike hypersurface.
For the linear momentum we get ($\xi = \partial _c$, $\varepsilon ^\beta
{}_\alpha =0$)
$$ \hor B _2 =  {1\over e} \left[ {\cal H} _c{}^{0a} -
   2 \delta ^a_{[c} \omega _{b] \alpha}{}^\beta
   {\cal H} ^\alpha {}_\beta {}^{0 b} \right] dS_a \quad , \eqa $$
for the angular momentum  ($\xi ^i $ given in (4.10)
$\varepsilon ^\beta {}_\alpha = g_{\alpha M}\delta _N^\beta -g_{N\alpha }
\delta _M ^\beta $)
$$ \hor B _2 =  {1\over e} \left[ 2 x_{[N}{\cal H}_{M]}{}^{0a} + 2 \Delta
   {\cal H}_{NM}{}^{0a} + 2 x_{[N} \omega _{M]\alpha }{}^\beta {\cal H}
   ^\alpha {}_\beta {}^{oa}\right] dS_a \quad ,\eqa $$
(observe, that we have for spatial rotations $\xi ^a dS_a=0$ ),
and for the boost ($\xi ^i $ given in (4.12),
$\varepsilon ^\beta {}_\alpha = g_{\alpha N}\delta _0^\beta -g_{0\alpha }
\delta _N ^\beta $)
$$ \hor B _2 =  {1\over e} \left[ 2 x_{[N}{\cal H}_{0]}{}^{0a} + 2 \Delta
   {\cal H}_{N0}{}^{0a} - 2 x_0 \delta ^a_{[b} \omega _{N]\alpha }{}^\beta
   {\cal H} ^\alpha {}_\beta {} ^{0b} - \omega _{b\alpha }{}^\beta {\cal H}
   ^\alpha {}_\beta {} ^{ba} \right] dS_a \quad .\eqa $$
Comparing the results with [6], we have to substitute ${\cal H}^{ij\alpha}
{}_\beta $ by twice of the momentum of Blagojevi\'c \& Vasili\'c because
their definition (of the generators of the Lorentz group and therefore) of the
rotational momentum differ by an factor 2 from our definitions.
Notice also that the leading term of the momentum is the Einstein part.
Finally recall that $\xo H_{\alpha \beta}$ gives no contribution to the
integrals, as shown in footnote (3). Then we can see that all of our
integrated expressions coincide with the ones of Blagojevi\'c \& Vasili\'c
(for the comparision of the integrals we have only to require (2.1)).

\medskip
At the end we want to make a comment on the expressions of [7] and [8].
In this articles are the field momenta $H_\alpha $ and $H_{\alpha \beta}$ used
as the integrands for total momentum and angular momentum, whereas the authors
considered also asymptotic anti de Sitter spacetimes.
As the field momenta bear indices, one has to choose carefully the
basis system (which is a tedious task for complicated configurations)
in order to get reasonable results. This was succesfully done in [7] for a
Schwarzschild -- anti-de-Sitter solution with torsion. But already the
application of this method on a Kerr -- anti-de-Sitter solution with torsion
leads to an infinite angular momentum as $H_{\alpha \beta}$ is proportional to
the curvature, and therefore does not have a suitable fall off.
Moreover, as it is obvious, these quantities do not give reasonable values
for solutions of the ECSK-theory or of GR.

\section{Conclusion}
We have discussed the behavior of the two expressions $B_1$, $B_2$ of eqs.
3.14,15 for conserved quantities (eq. 3.16) of PGT in asymptotic flat
spacetimes.
We have seen that the variation of the accompanying Hamiltonians are well
defined for appropopriate phase spaces. The respective phase spaces of the
two expressions differ slightly, whereas the appropopriate  phase space of
$B_2$ (see eqs. 4.11,18) is larger than the one of $B_2$ (see eq. 5.9).
Finally we have seen that the expression $B_2$, for appropriate boundary
conditions, coincides with those of Hayashi \& Shirafuji and Blagojevi\'c \&
Vasili\'c.
In [4] both expressions were tested with asymptotic flat and asymptotic
constant curvature solutions of PGT and gave, for the tested solutions,
the same results as the corresponding solutions of General Relativity.
The advantage of the expressions is that they  are not restricted
on an asymptotically Cartesian basis and that they can be also used
in asymptotic anti-de-Sitter spacetimes. Moreover it gives one compact
expression for all of the ten conserved quantities.

\section{Acknowledgement}
The author is grateful to Prof. J. M. Nester for his continous support and
useful discussions and to Prof. F. W. Hehl for many usefull hints.
This work was supported by the National Science Council of the Republic of
China under contract no. NSC83-0208-M-008-014.

\section{Literature}
\newref  T. Regge and C. Teitelboim, Ann. Phys. {\bf 88} (1974) 286
\newref  J. M. Nester, Mod. Phys. Lett. A {\bf 6} (1991) 2655
\newref  R. D. Hecht, ``Erhaltungsgr\"o\ss en in der Poincar\'e-%
             Eichtheorie der Gravitation'', Ph. D. thesis, Univ. of Cologne,
             1993
\newref  R. D. Hecht and J. M. Nester, Phys. Lett. {\bf A180}, (1993), 324
\newref  K. Hayashi and T. Shirafuji,
                Prog. Theor. Phys. {\bf 73 } (1985) 54
\newref  M. Blagojevi\'c and M. Vasili\'c, Class. Quantum Grav. {\bf 5}
             (1988) 1241
\newref  P. Baekler, R. Hecht, F. W. Hehl and T. Shirafuji, Prog. Theor.
         Phys. {\bf 78} (1987) 16
\newref  E. W. Mielke and R. P Wallner, Nuovo Cimento {\bf 101B} (1988) 607;
         {\bf 102B} (1988) 555
\newref  K. Hayashi and T. Shirafuji, Prog. Theor. Phys. {\bf 64} (1980)
         866, 883, 1435, 2222
\newref  F. W. Hehl, ``{\it Four Lectures on Poincar\'e gauge field
                theory}'', in : Proceedings of the 6th Course of the
                International School of Cosmology and Gravitation on
                Spin,  Torsion and Supergravity, P.G.Bergmann and
                V.de Sabatta eds.,(Plenum Press, New York, 1980) p 5
\newref  J. D. McCrea, Class. Quantum Grav. {\bf 9} (1992) 553
\newref  H.H. Chen, D.C. CherH.H. Chen, D.C. Chern, R.R. Hsu, J.M. Nester and
         W.B. Yeung, Prog. Theor. Phys. {\bf 79} (1988) 77
\newref  J. Nester ``Lectures on gravitational gauge theory'', Nat. Tsing-Hua
         University, Hsin-chu, Taiwan (1989), not published
\newref  E. Nahmad-Achar and B. F. Schutz, Class. Quantum Grav. {\bf 4}
         (1987) 929; Gen. Rel. Grav. {\bf 19} (1987) 655
\newref  L. D. Landau and E. M. Lifschitz, ``The Classical Theory of Fields''
         2nd ed. (Addison-Wesley, Reading, Mass.,1962)
\newref  C. M\o ller, Mat. Fys. Skr. Dan. Vid. Selsk. {\bf 1} (1961);
         Mat. Fys. Skr. Dan. Vid. Selsk. {\bf 35} (1966)

\bye